\documentclass[epj]{webofc}
\pdfoutput=1
\usepackage[utf8]{inputenc}
\usepackage[varg]{txfonts}   
\usepackage{booktabs}
\usepackage{slashed}
\usepackage{xcolor}
\definecolor{darkred}{rgb}{0.4,0.0,0.0}
\definecolor{darkgreen}{rgb}{0.0,0.4,0.0}
\definecolor{darkblue}{rgb}{0.0,0.0,0.4}
\usepackage[bookmarks,linktocpage,colorlinks,
    linkcolor = darkred,
    urlcolor  = darkblue,
    citecolor = darkgreen]{hyperref}
\usepackage{subfigure}
\wocname{EPJ Web of Conferences}
\woctitle{Lattice2017}
\def\Eq#1{Eq.~(\ref{#1})}
\def\sss{\scriptscriptstyle}

\def\argphi{\,\mathrm{Arg}\,\varphi}
\def\Tchop{T_{\mathrm{chop}}}
\def\tA{\theta_{\mathrm{a}}}
\def\rhodm{\rho_{\mathrm{dm}}}

\def\rmin{r_{\mathrm{min}}}

\def\nax{n_{\mathrm{ax}}}

\def\half{\frac{1}{2}}

\def\LL{\mathcal{L}}
\def\DD{\mathcal{D}}
\def\Tr{\,\mathrm{Tr}\,}
\def\be{\begin{equation}}
\def\ee{\end{equation}}
\def\bea{\begin{eqnarray}}
\def\eea{\end{eqnarray}}

\begin{document}
%
\selectlanguage{english}
\title{%
Axion dark matter and the Lattice
}
\author{%
\firstname{Guy} \lastname{Moore}\inst{1}\fnsep\thanks{Acknowledges
  financial support by TU Darmstadt}
}
\institute{%
Institut f\"ur Kernphysik, Technische Universit\"at Darmstadt,
Schlossgartenstra{\ss}e 2, D-64289 Darmstadt, Germany
}
\abstract{%
First I will review the QCD theta problem and the Peccei-Quinn
solution, with its new particle, the axion.  I will review the
possibility of the axion as dark matter.  If PQ symmetry was restored
at some point in the hot early Universe, it should be possible to make
a definite prediction for the axion mass if it constitutes the Dark
Matter.  I will describe progress on one issue needed to make this
prediction -- the dynamics of axionic string-wall networks and how
they produce axions.  Then I will discuss the sensitivity of the
calculation to the high temperature QCD topological susceptibility.
My emphasis is on what temperature range is important, and what level
of precision is needed.
}
\maketitle
\section{Overview}\label{intro}

The axion
\cite{Peccei:1977hh,Peccei:1977ur,Weinberg:1977ma,Wilczek:1977pj}
is a proposed particle, the angular excitation of a
new ``Peccei-Quinn'' (PQ) field $\varphi$ that would solve the strong
CP problem \cite{tHooft:1976,Jackiw:1976pf,Callan:1979bg} and which is
also a very interesting dark matter candidate
\cite{Preskill:1982cy,Abbott:1982af,Dine:1982ah}, thereby solving
two puzzles with one mechanism.  That's why I think it well motivated
to study the axion as a dark matter candidate.  The axion
model has one undetermined parameter, the vacuum value of $\varphi$,
$f_a$; the axion mass $m_a$ scales as $f_a^{-1}$.  The value of $f_a$
also plays a role in determining the amount of axion dark matter
produced in the early Universe.  So do some nontrivial dynamics which
we will explain in detail below.  If we can understand the nontrivial
dynamics of the axion field during cosmology, that lets us find a
fixed relation between $f_a$ and the (measured) dark matter
abundance.  It therefore allows a clean determination of the axion
mass, under some simple assumptions.  This is valuable in the
experimental search for the axion and it motivates us to solve the
cosmological axion dynamics.

Supposing we make the following assumptions:
\begin{itemize}
\item
  The axion exists.
\item
  The axion field starts out ``random'' (in a sense we will define
  precisely below) either during or shortly after inflation (or
  whatever physics featured in the Universe at very high energy
  density).
\item
  Gravity and the Universe's energy budget followed the ``standard''
  picture (General Relativity and the known Standard Model species
  dominating the energy density) at temperatures $T < 2$ GeV.
\item
  Axions make up all of the observed dark matter.
\end{itemize}
Then, as I will explain, we have enough information to determine the
axion mass.  But to do so we will need to solve two problems:
\begin{enumerate}
\item \label{item1}
  We need to know the temperature dependence of the QCD topological
  susceptibility at temperatures between 540 and 1150 MeV.
\item \label{item2}
  We need to control the axion field dynamics during this temperature
  range, to solve for the efficiency of axion production.
\end{enumerate}
The session following this talk addresses item \ref{item1} and the
remainder of this talk will lay out the groundwork more completely and
will then address item \ref{item2}.

\section{Strong CP problem and the axion}

Let me refresh your memories on the strong CP problem.
There are two gauge-invariant dimension-4 scalars which can enter the
gauge-field part of the QCD Lagrangian:
\begin{equation}
  \label{QCDLagrangian}
  \LL = \frac{1}{2g^2} \Tr G_{\mu\nu} G_{\mu\nu}
  + \frac{\Theta}{32\pi^2} \Tr \epsilon_{\mu\nu\alpha\beta}
  G^{\mu\nu} G^{\alpha\beta} \,.
\end{equation}
The latter term is P and T odd because it contains the antisymmetric
tensor $\epsilon_{\mu\nu\alpha\beta}$.  The operator which $\Theta$
multiplies is the topological density and it integrates to the
instanton number.  Its zero-momentum two-point function defines the
topological susceptibility,
\begin{equation}
  \label{chidef}
  \chi(T) \equiv \left\langle \int d^4 x
  \frac{1}{32\pi^2} \epsilon_{\mu\nu\alpha\beta} \Tr G_{\mu\nu}
  G_{\alpha\beta}(x) \;
  \frac{1}{32\pi^2} \epsilon_{\sigma\rho\kappa\zeta}
  \Tr G_{\sigma\rho} G_{\kappa\zeta}(0) \right\rangle_{T} \,,
\end{equation}
$\langle \ldots \rangle_T$ means the expectation value in the thermal
ensemble at temperature $T$.

Such a term is strongly constrained by the absence of a measured
neutron electric dipole moment.  The experimental limit of
\cite{Baker:2006ts}
\begin{equation}
  |d_{n,\mathrm{meas}}| < 2.9\times 10^{-26} \; e\,\mathrm{cm}
\end{equation}
contradicts the lattice results for the dipole moment from $\Theta$
\cite{Guo:2015tla}
\begin{equation}
  d_n = -3.8 \times 10^{-16} \; e\,\mathrm{cm} \times \Theta
\end{equation}
unless $|\Theta|<10^{-10}$.  At this conference we saw new results for
the lattice $\Theta$-dependent dipole moment which show that the above
result may be too high and its error bar was certainly underestimated
(see \cite{Abramczyk:2017oxr,Liu:2017man} and these proceedings).
However it is clear that the absence of a neutron electric dipole
moment places an extremely tight constraint on $\Theta$.

This is hard to understand because we know P and T are not fundamental
symmetries; and any physics at a high scale which violates them generically
gives rise to a $\Theta$ which does not decrease as we move to lower
scales.  For instance, consider a very heavy Dirac quark species $Q$,
with Lagrangian
\begin{equation}
  \label{Qlagrangian}
  \LL_Q = \bar{Q} \slashed{D} Q +
  \left( m \bar{Q} P_{\sss L} Q + m^* \bar{Q} P_{\sss R} Q \right) \,.
\end{equation}
Note that $m$ can be complex, since
$(\bar{Q} P_{\sss L} Q)^\dagger = \bar{Q} P_{\sss R} Q$.  But an
imaginary part is T and P odd, since the role of left and right
projector, $P_{\sss L}$ and $P_{\sss R}$, switch under parity and
because T is antiunitary.  We can remove this mass through a chiral
rotation of $Q$,
\begin{equation}
  \label{Qtransform}
  Q \to (e^{-i\,\mathrm{arg}\, m/2} P_{\sss L} +
  e^{-i\,\mathrm{arg}\, m/2} P_{\sss R}) Q \,,
\end{equation}
at the cost of reintroducing it, via the Fujikawa mechanism
\cite{Fujikawa:1979ay,Fujikawa:1980eg}, as a shift in the value of
$\Theta$,
\begin{equation}
  \label{Fujikawa}
  \Theta \to \Theta - \mathrm{arg}\, m \,.
\end{equation}
Therefore even a very heavy quark can influence the P and T symmetry
properties of low energy QCD.

But what if there is a symmetry forbidding the mass term for this
quark?  For instance, suppose $P_{\sss L} Q$ is charge 1 and
$P_{\sss R} Q$ is charge-0 under some global U(1) symmetry?  Then the
mass term breaks this symmetry, but a complex scalar $\varphi$ with
charge 1 under the symmetry could induce a mass via a Yukawa
interaction and a vacuum value.  The possible Lagrangian terms for
such a scalar are
\begin{equation}
  \label{Laxion1}
  \LL_\varphi = \partial_\mu \varphi^* \partial_\mu \varphi
  + \frac{m^2}{8f_a^2} \left( f_a^2 - 2\varphi^* \varphi \right)^2
  + \Big( y \varphi \bar{Q} P_{\sss R} Q
  + y^* \varphi^* \bar{Q} P_{\sss L} Q \Big) \,.
\end{equation}
The combination $y \varphi$ plays the role of $m^*$ in the previous
case.  But now $\argphi$ is a dynamical quantity.  We will be
interested in temperatures around 1 GeV and $\varphi$ varying on
scales of the Hubble scale at that time -- tens of meters!  Therefore
from the point of view of QCD we can take $\varphi$ to be
space-independent, and perform an $\argphi$ dependent rotation on
$Q$, making the theta term
\begin{equation}
  \Theta \to \Theta + \argphi \,.
\end{equation}
Here we have absorbed the phase in $y$ into a phase redefinition of
$\varphi$.  We can also absorb $\Theta$ in the same way, so that
$\argphi$ alone plays the role of $\Theta$-angle.
For notational compactness we will henceforth write
$\varphi = v e^{i\tA}$, with $\tA = \argphi$.  This specific way of
coupling QCD topology to a complex scalar is called the KSVZ axion
\cite{Kim:1979if,Shifman:1979if}.  There are other mechanisms but the
low-energy phenomenology is essentially identical and this mechanism
is particularly clear to understand.

From the point of view of QCD, the $\Theta$-angle is replaced by a
possibly spacetime-varying dynamical field $\tA$.  What about from the
point of view of the field $\varphi$?  Since we want physics on the
meter length scale, we can integrate out QCD, leading to an effective
potential:
\begin{eqnarray}
  \label{Veff}
  V_{\mathrm{eff}}(\tA) & = & -\frac{T}{\Omega} \ln
  \int \DD (A_\mu \bar\psi \psi)\, \mathrm{Det}\,(\slashed{D}{+}m)
  e^{-\int d^4 x \frac{1}{2g^2} \Tr G_{\mu\nu} G_{\mu\nu}}
  \times e^{i\tA \int d^4 x \frac{1}{32\pi^2}
    \epsilon_{\mu\nu\alpha\beta} \Tr G_{\mu\nu} G_{\alpha\beta}}
  \nonumber \\
  & \simeq & \chi(T) ( 1 - \cos\,\tA ) \,,
\end{eqnarray}
with $\Omega$ the volume of space included in the path integration.
In the second line we have made a dilute instanton approximation,
which is that the integration exponentiates over the two-point
function of the topological density, controlled by the topological
susceptibility $\chi(T)$ introduced already in \Eq{chidef}.
This is not a good approximation for large $\tA$ and low temperatures
\cite{diCortona:2015ldu}, but it works well when instantons are
dilute, which is true for $T > 500\: \mathrm{MeV}$, and for small
values of $\tA$, which will be all we encounter below this
temperature.  So we can actually use this approximation all the time.
Independent of this approximation, it is easy to see that the
effective potential is smallest (the $\tA$ choice is most
energetically favored) for $\tA=0$ and therefore when P and T symmetry
are restored.  Note that $\tA$ only enters as the coefficient in a
complex phase, in an otherwise real and positive integral.  The
integral is maximized, and the free energy minimized, if the phase is
always unity.  Any nonzero value of $\tA$ gives rise to phase
cancellations and therefore suppresses the partition function, raising
the free energy.

Although we derived it in Euclidean space, we can also use this
effective potential in Minkowski space to study the spacetime
evolution of the $\varphi$ field.
In summary, the Minkowski effective Lagrangian for the $\varphi$ field
is
\begin{equation}
  \label{Laxion2}
  -\LL_\varphi = \partial_\mu \varphi^* \partial^\mu \varphi
  + \frac{m^2}{8f_a^2} \left( f_a^2 - 2\varphi^* \varphi \right)^2
  + \chi(T) ( 1 - \cos\tA ) \,.
\end{equation}
We will use this to determine the dynamics of the field in the next
sections.

\section{Axion in Cosmology}

Let us see what happens to the axion field during cosmological
evolution.

\subsection{Value of susceptibility}

\begin{figure}[thb] 
  \centering
  \includegraphics[width=7.5cm,clip]{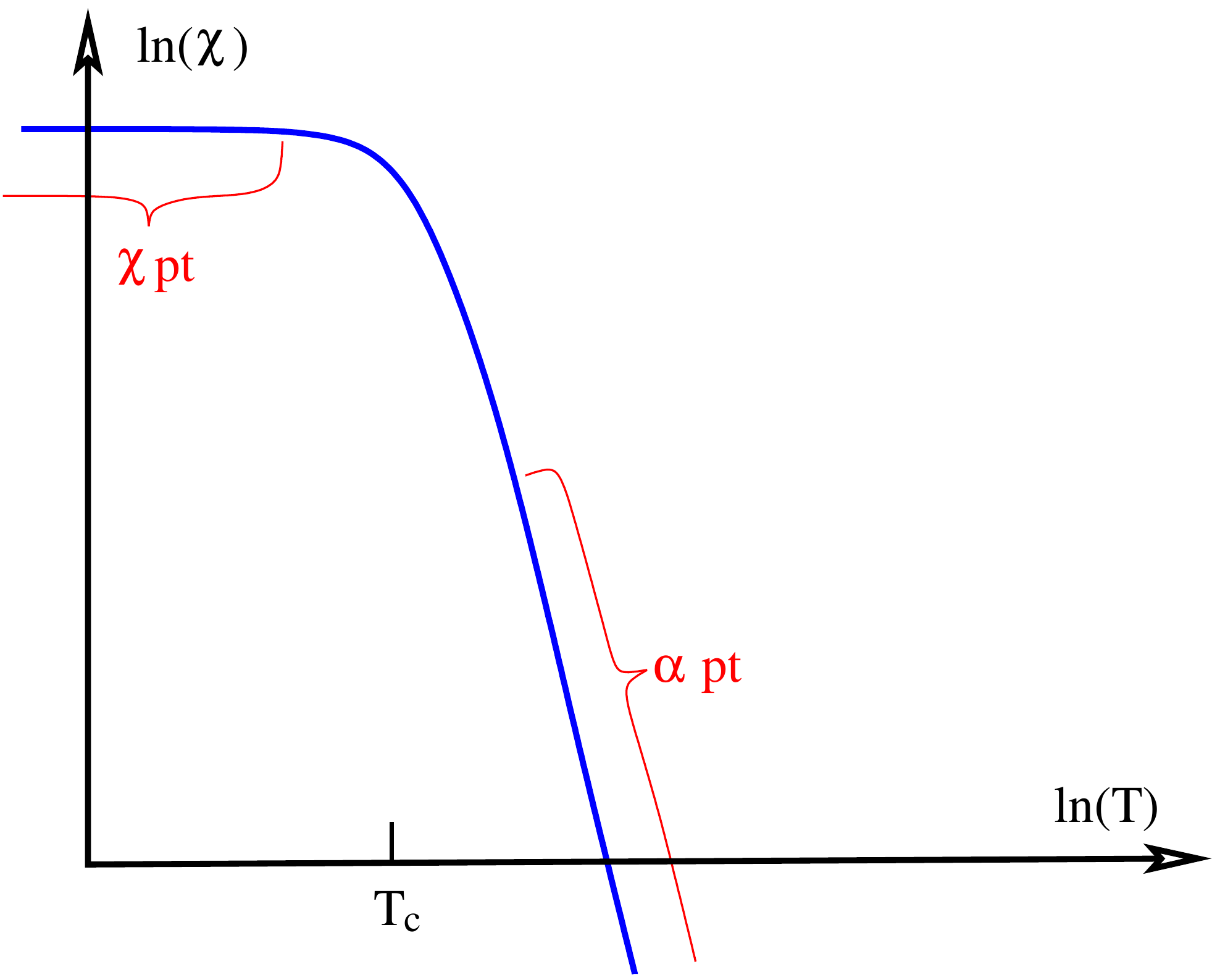}
  \caption{Cartoon of what we know about the topological
    susceptibility as a function of temperature (log-log plot).
    At low temperatures, chiral perturbation theory gives us the
    zero-temperature limit and small-temperature behavior.  At large
    temperatures, perturbation theory predicts a power-law falloff
    with a power near 8.}
  \label{fig:chicartoon}
\end{figure}

The form of \Eq{Laxion2} makes it clear that, in order to study the
axion's role in cosmology, we are going to need to know the
temperature dependence of the topological susceptibility $\chi(T)$.
It does not yet tell us what temperature range will be interesting.
Figure \ref{fig:chicartoon} shows our knowledge at the cartoon level.
At low temperature or vacuum, chiral perturbation theory works and
\cite{diCortona:2015ldu}
\begin{equation}
  \label{Chi0}
  \chi(T \to 0) \simeq \frac{m_u m_d}{(m_u+m_d)^2} m_\pi^2 f_\pi^2
  \simeq (76 \:\mathrm{MeV})^4 \,.
\end{equation}
At high temperatures we have conventional perturbation theory,
which forecasts \cite{Gross:1980br} that
$\chi(T) \propto T^{-7-N_f/3}$.  However the exact coefficient is
sensitive to the physics of electric screening and is not known
accurately. This is why we need lattice results for this quantity!
Recently there have been several
\cite{Berkowitz:2015aua,Borsanyi:2015cka,Petreczky:2016vrs,
  Taniguchi:2016tjc,Burger:2017xkz,Frison:2016vuc,Borsanyi:2016ksw},
which give generally compatible results but generally at temperatures
below 600 MeV (or in the quenched approximation).  It takes new
techniques to reach higher temperatures, and only one recent paper
\cite{Borsanyi:2016ksw} achieves this, reaching temperatures of 1500
MeV.  Also, one group
\cite{Bonati:2015vqz,Bonati:2016imp} finds results which are
discrepant with the others, indicating that the matter is not yet
settled.  Here we will assume that the results of
\cite{Borsanyi:2016ksw} are correct.  This may well be the case, but
we leave the discussion of the relative merits of these approaches and
results to the panel, who have more expertise.
Needless to say it would be valuable to know definitively that
$\chi(T)$ is well determined.

\subsection{Space-uniform axion field}

So let's assume for now that we know $\chi(T)$.  For simplicity let us
also assume that the axion takes the same value everywhere in space,
$\tA(x,t) = \tA(t)$.  It is simplest to work in terms of conformal
time, so the metric is $g_{\mu\nu} = a^2(t) \eta_{\mu\nu}$ with $a$
the scale factor. (Later we will use $a$ to represent the lattice
spacing.  This is actually the same thing, since we will work in
comoving coordinates; the lattice spacing is proportional to the scale
factor and we may as well use a proportionality of 1.)
In the radiation era $a(t) \propto t$ and $T \propto t^{-1}$.  The
radial component of $\varphi = v e^{i\tA}$ is inactive, $v=f_a$, and
the angular part obeys
\begin{eqnarray}
  \label{Lax3}
  \LL & = & f_a^2 t^2 \left( \frac 12 (\partial_t \tA)^2
  + t^2 \tilde\chi(t^{-1}) (1 - \cos \tA) \right) \,,
  \\
  \label{EOMax}
  \partial_t^2 \tA + \frac{2}{t} \partial_t \tA & = &
  - t^2 \tilde\chi(t^{-1}) \sin \tA \,,
\end{eqnarray}
where $\tilde\chi(t^{-1})$ is a rescaled form of the susceptibility.
This leads to damped, anharmonic oscillations.  The oscillations start
roughly at the time $t_*$ when $t_*^2 \tilde\chi(t_*^{-1}) = t_*^{-2}$, or
in physical units, when $m_a \equiv \sqrt{\chi(T)/f_a^2}$
obeys $m_a t_* = 1$ or equivalently $m_a / H=1$ with $H$ the Hubble
scale.  After this time the oscillations accelerate as
$t^2 \tilde\chi$ increases, and they damp away.  The damping arises
both from the $2\partial_t \tA/t$ term (Hubble drag) and from the time
variation of the susceptibility.  After several oscillations the axion
particle number becomes an approximate adiabatic invariant, with
number density parametrically of form $(t_*/t)^2 (f_a^2/t_*)$.
We see that the number density is quadratic in $f_a$, while the axion
mass is $m_a \propto f_a^{-1}$.  Because $\chi(T)$ is a very strong
function of temperature, $t_*$ depends only weakly on $f_a$, and so
the generated axion energy density is almost linear in $f_a$.

Therefore, the larger the value of $f_a$, the larger the produced
axion abundance.  However the axion abundance also depends on the
unknown initial angle $\tA(t=0)$.  Therefore the dark matter density
depends on two variables and it is impossible to make a clean
prediction for the value of $f_a$.
We can make a baseline prediction, however, by averaging over the
value of the starting angle $\tA(t=0)$.  Doing so, one finds
the axion mass should be $32\, \mu\mathrm{eV}$, and $t_*$ corresponds
to a temperature of $T_* = 1.6\,\mathrm{GeV}$.

\subsection{Space-random axion field}

It is far more likely that the Universe started out with a spatially
random value for $\tA$, with no correlations on scales longer than the
Hubble scale.  Arguments for this picture are presented in
\cite{Visinelli:2014twa} and are summarized as follows:
\begin{itemize}
\item
  It is likely that inflation occurs with a high scale,
  $H^2 > f_a^2 / 60$.  In this case, over 60 efoldings of inflation,
  quantum fluctuations stretched (squeezed) by inflation into
  classical fluctuations would randomize the value of the axion field
  over the course of inflation.  The observation of cosmological
  tensor modes would more-or-less settle this issue.
\item
  After inflation, the Universe reheats to a temperature which can be
  as high as $T_{\mathrm{rh}} \sim 0.1 \sqrt{Hm_{\mathrm{pl}}}$.
  Even if $H \ll f_a$, if the reheat temperature is
  $T_{\mathrm{rh}} > f_a$, there would be thermal symmetry
  restoration for $\varphi$.  Then when the temperature falls below
  this scale, $\varphi$ would independently take on a vacuum value at
  different points in space, which would be uncorrelated.
\item
  The case where inflation and reheating are both low-scale
  is actually tightly constrained by the absence of observed
  isocurvature fluctuations (different fluctuations in $\tA$ than in
  the radiation temperature), which require roughly
  $H < 10^{-5} f_a$.  Most inflation model-builders would consider
  this rather unlikely.
\end{itemize}
I emphasize that we do not \textsl{know} that $\tA$ was randomized
in the early Universe (assuming the axion exists).  But it appears
likely, and it motivates studying the consequences.
I will also assume that the axion makes up all of the dark matter in
the universe, so we may equate the final axion matter density with the
dark matter density, which is known to obey
$\rhodm/s = 0.39 \: \mathrm{eV}$ with $s$ the entropy density
\cite{Ade:2015xua}.

The space-inhomogeneous case is much more complicated than the
space-homogeneous case.  Nevertheless, in the remainder of the
presentation I will show how to solve it.

\section{Axion string/wall network}

The $\varphi$ field varies with amplitude of order
$f_a \sim 10^{11}\,\mathrm{GeV}$ over a length scale controlled by
$H \sim T^2/m_{\mathrm{pl}} \sim 10^{-18} \,\mathrm{GeV}$.  This huge
hierarchy makes the dynamics those of a classical field to extremely
high accuracy.  The Lagrangian \Eq{Laxion2} (times $t^2$ to account
for Hubble expansion) and resulting classical
equations of motion are easy to put on the lattice and
solve as a function of time, from random initial conditions.  In broad
brushstrokes, our approach is to do just this, evolving the system
until only small fluctuations in $\tA$ remain and their evolution has
become adiabatic.  Then we integrate the associated axion number,
\begin{equation}
  \label{nax}
  \nax = \int \frac{d^3 p}{(2\pi)^3} f(p)
  = \int \frac{d^3 p}{(2\pi)^3}
  \frac{(p^2+m_a^2) \varphi^* \varphi(p) + \dot\varphi^* \dot\varphi}
       {\sqrt{p^2+m_a^2}}
\end{equation}
and compare it to the result of the angle-averaged misalignment
baseline.

In fact such a simulation is not sufficient, because of the large
hierarchy in \Eq{Laxion2} between the mass scale
$m \sim 10^{11}\,\mathrm{GeV}$ of radial
excitations and the mass scale
$m_a = \sqrt{\chi(T)}/f_a \sim H \sim 10^{-18}\,\mathrm{GeV}$ of
angular fluctuations.  The simulations have to take place at the
$m_a$ scale, which means that the radial-mass scale cannot be
resolved.  Naively this should not matter, as radial excitations
should decouple.  But it does matter, because the theory contains
topological string defects which play a role in the dynamics, and the
string tension depends logarithmically on the ratio $m/m_a$.
Let's explain this in a little more detail.

\subsection{String defects}

First note that $\tA$ is only defined modulo $2\pi$.  Therefore in
traversing a circle, $\tA$ might return to its starting value, but it
might only return modulo $2\pi$, that is,
$\oint \partial_i \tA dx^i = 2\pi N$.  The integer $N$ is a winding
number which counts a ``flux'' of string defects through the circle.
If we deform a loop, $N$ can only change when the loop passes through
a singularity in the $\tA$ field.  The locus of these singularities
defines the axionic cosmic string.

\begin{figure}[thb] 
  \centering
  \includegraphics[width=9cm,clip]{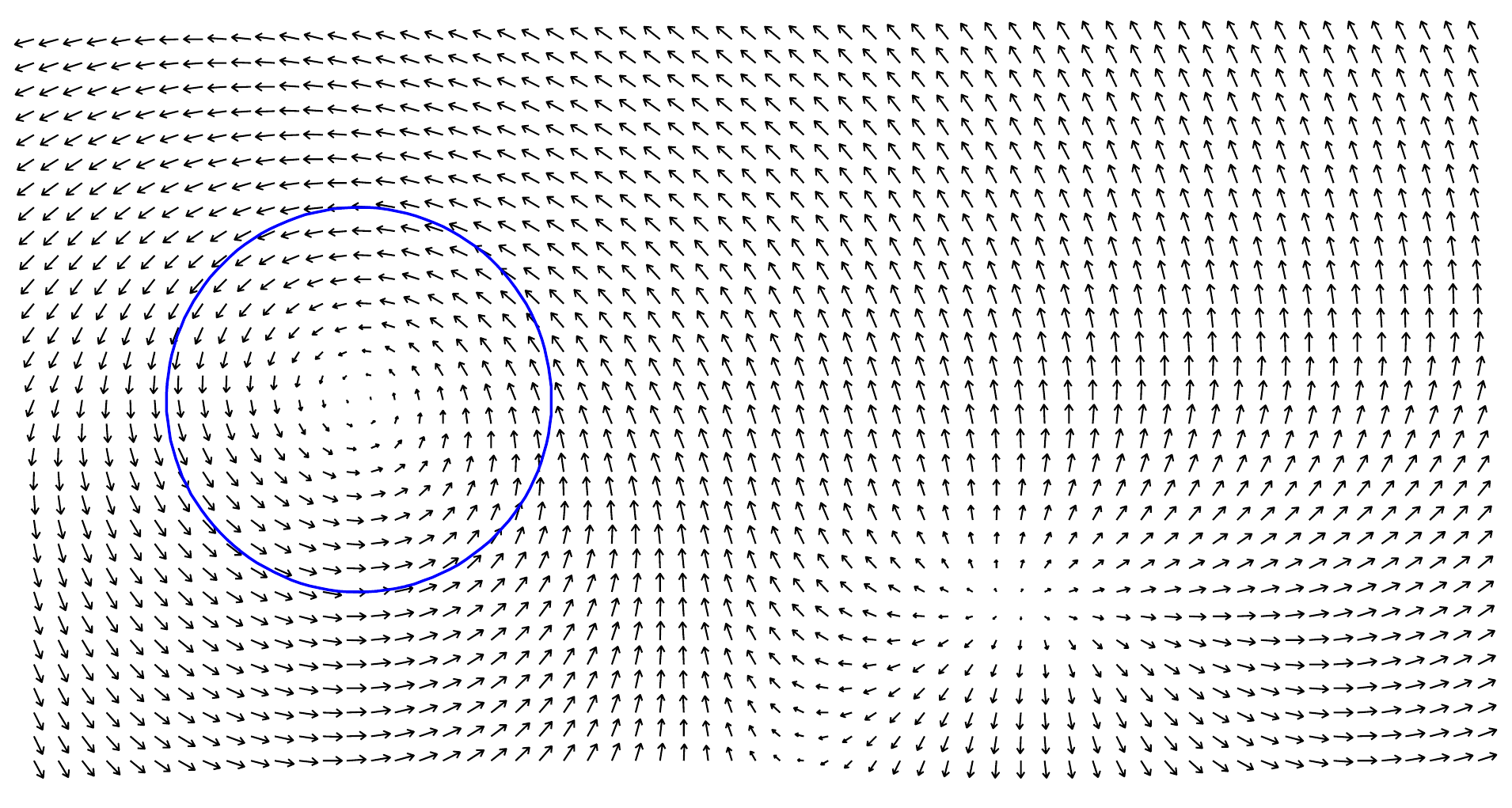}
  \caption{A 2D slice of a simulation, with the complex field
    $\varphi$ represented by a field of arrows with a length and
    direction.  Going around the blue circle, the arrow direction
    winds by $2\pi$.  The associated string defect is at the center of
    the circle; another string defect is farther down and to the
    right.}
  \label{fig:string}
\end{figure}

We illustrate the idea with Figure \ref{fig:string}, which shows a 2D
slice out of a simulation, representing the complex field as a field
of arrows with length and direction.  The field direction has
singularities where the arrows have zero length; going around the
singularity, the direction of $\varphi$ revolves by $\pm 2\pi$.  The
singular point extends in 3D into a line where the field has zero
value; any loop circling this line will have the direction of
$\varphi$ revolve by $\pm 2\pi$ as the one circles around the string.

Such a defect -- essentially a vortex in the $\varphi$ field -- is
called an axionic
cosmic string, and it is topologically stable; no local changes to the
value of $\varphi$ can cause it to disappear.  If PQ symmetry is
restored in the early Universe, then $\tA$ starts out uncorrelated at
widely separated points and will generically begin with a dense
network of these strings (the Kibble mechanism for string production
\cite{Kibble:1976sj}).  The strings evolve, straightening out, chopping off
loops, and otherwise reducing their density, arriving at a scaling
solution \cite{Albrecht:1984xv} where the length of string per unit 
volume scales with time t as $t^{-2}$.  They may play a dominant role
in establishing axion production in the scenario under discussion
\cite{Davis:1986xc}.

Let us analyze the structure of a string in a little more detail.
Consider a straight string along the $z$ axis; in polar $(z,r,\phi)$
coordinates the string equations of motion are solved by
$\sqrt{2} \varphi = v(r) f_a e^{i\phi}$, with $v(r) \simeq 1$ for all
$r \gg 1/m$; so $\tA = \phi$ (up to a constant which
we can remove by our choice of $x$-axis).  The string's energy is
dominated by the gradient energy due to the space variation of $\tA$:
\bea
\label{Tension}
T_{\mathrm{str}} &=& \frac{\mbox{Energy}}{\mbox{length}}
= \int r\,dr\,d\phi \left( V(\varphi^* \varphi) + \half \nabla \varphi^*
\nabla \varphi \right)
 \\
\label{kappa}
 & \simeq & \pi \int r\, dr \left( \frac{\partial_\phi \varphi^*}{r}\;
\frac{\partial_\phi \varphi}{r} \right) \simeq
\pi \int^{H^{-1}}_{1/m} r \, dr \; \frac{f_a^2}{r^2}
= \pi f_a^2 \ln(m/H) \equiv \pi f_a^2 \kappa \,,
\eea
where the integral over $r$ is cut off at small $r$ by the scale where
$v(r) \neq 1$ (the string core), and at large distances by the scale
where the string is not alone in the Universe but its field is
modified by other strings or effects; this should be the larger of $H$
and $m_a$.  We define
$\kappa = \ln(m/H)$ as the log of this scale ratio.  Now $m$ is at
most $f_a \sim 10^{11}\,\mathrm{GeV}$, and to ensure that the radial
particles decay by the scale of 1 GeV we need $m >
10^3\,\mathrm{GeV}$.  Therefore $\kappa \in [48,67]$.

This logarithm, $\kappa$, controls several aspects of the strings'
dynamics.  It controls the string tension, as we just saw.  More
relevant, while the string tension is $\pi \kappa f_a^2$, the
string's interactions with the long-range $\varphi$ field scale as
$f_a^2$ \textsl{without} the $\kappa$ factor.  Therefore the string's
long-range interactions become less
important, relative to the string evolution under tension, as $\kappa$
gets larger.  The long-range interactions are responsible for energy
radiation from the strings, as well as for long-range, often
attractive, interactions between strings.  Since these effects tend to
deplete and straighten out the string network, the large-$\kappa$ theory
will have denser, kinkier strings.  Indeed, in the large $\kappa$
limit the string behavior should go over to that of local (Nambu-Goto)
strings \cite{Dabholkar:1989ju}.
Unfortunately, a numerical implementation must resolve the length
scale $m$, $ma \leq 1$, and cannot exceed $m/H \sim 1000$; numerical
studies of the scalar field system have $\kappa < 7$, nearly an order
of magnitude too small.

\subsection{Wall defects}

\begin{figure}[thb] 
  \centering
  \includegraphics[width=9cm,clip]{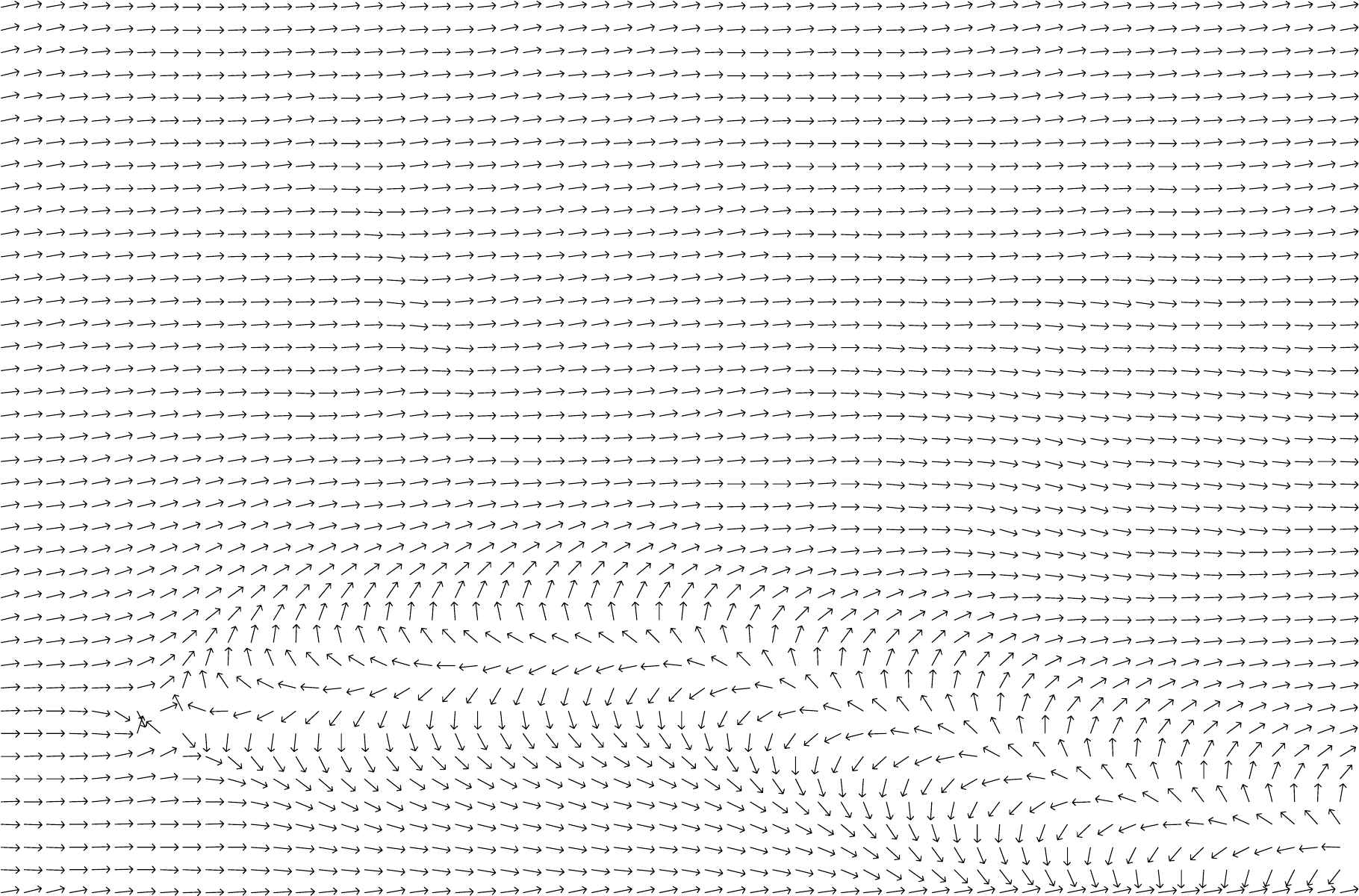}
  \caption{A 2D slice of a simulation, this time with a string
    (lower left) attached to a domain wall which extends near the
    bottom, off the right edge.}
  \label{fig:wall}
\end{figure}

Besides the strings, there are also wall defects.  These occur late in
the simulation when $m_a \gg H$.  The potential term
$\chi(T) (1-\cos\tA)$ then forces $\tA \simeq 0$
nearly everywhere -- modulo $2\pi$.  But suppose some region has
$\tA \simeq 0$ and another has $\tA \simeq 2\pi$.  There must be some
2D surface between them with $\tA \simeq \pi$.  This is a wall
defect.  The region near the defect where $\tA$ differs significantly
away from its minimum has thickness $\ell \sim 1/m_a$, which is easily
resolved on the lattice.  The surface tension of such a surface turns
out to be $8 m_a f_a^2$.  A 2D slice of a configuration, illustrating
such a domain wall attached to a string, is shown in Figure
\ref{fig:wall}.

These wall defects are not a problem to
simulate.  But they play an important role in the dynamics.  Every
string has $\tA$ take every value $[0,2\pi]$ as one goes around the
string.  That includes $\tA=\pi$.  Therefore every string is attached
to a domain wall.  When $m_a$ becomes $m_a \gg \kappa H$, the force
from the domain wall tension becomes large enough to pull around the
strings, leading to the collapse of the string network and the
annihilation of all strings.  It is only after this network collapse
that one can speak about axion number.  Because of the factor $\kappa$
in the needed tension, a large-$\kappa$ simulation will feature a more
persistent string network.

One final problem for scalar-only simulations, pointed out in
\cite{axion1}, is that the domain walls actually lose even their
metastability as soon as $m_a^2/m^2 > 1/39$.  This drives up the
required size of simulations so that large $m$ can be achieved.

\section{Simulating high-tension strings}

We see from the previous section that simulations of the $\varphi$
field alone are not reliable.  Although one can make the scale $m$
very heavy compared to $H,m_a$, the string tension depends
logarithmically on this scale, and is nearly a factor of 10 too
small.  This profoundly affects the dynamics of the string network,
and therefore renders the results unreliable.  We need a method to
simulate high-tension strings coupled to $\tA$.
We found such a method in \cite{axion3} and present it here.

\subsection{Effective theory}

We are interested in the large-scale structure of string
networks and the infrared behavior of any (pseudo)Goldstone modes they
radiate.  For these purposes it is
not necessary to keep track of all physics down to the scale of the
string core.  Rather, it is sufficient to describe the desired IR behavior
with an \textsl{effective theory} of the strings and the Goldstone
modes around them.  This consists of replacing the physics very close
to the string core with an equivalent set of physics.  It has long
been known how to do this \cite{Dabholkar:1989ju}.  The string cores
are described by the Nambu-Goto action
\cite{Goto:1971ce,Goddard:1973qh,Nambu:1974zg}, which describes the
physics generated by the string tension arising close to the string
core.  The physics of the Goldstone mode is described by a Lagrangian
containing the scalar field's phase.  And they are coupled by the
Kalb-Ramond action \cite{Kalb:1974yc,Vilenkin:1986ku}:
\begin{align}
  \label{Ltot}
  \LL & = \LL_{\mathrm{NG}} + \LL_{\mathrm{GS}} + \LL_{\mathrm{KR}} \,,
  \\
  \label{LNG}
  \LL_{\mathrm{NG}} & =  \bar\kappa \pi f_a^2 \int d\sigma
  \sqrt{{y'}^2(\sigma)(1-\dot{y}^2(\sigma))} \,,
  \\
  \label{LGS}
  \LL_{\mathrm{GS}} & = f_a^2 \int d^3 x \; \partial_\mu \tA \partial^\mu \tA \,,
  \\
  \label{LKR}
  \LL_{\mathrm{KR}} & = \int d^3 x \; A_{\mu\nu} j^{\,\mu\nu} \,,
  \\
  \label{Hmna}
  H_{\mu\nu\alpha} & = f_a \epsilon_{\mu\nu\alpha\beta}
  \partial^\beta \tA = \partial_\mu A_{\nu\alpha} + \mbox{cyclic} \,,
  \\
  \label{jmn}
  j^{\,\mu\nu} & = -2\pi f_a \int d\sigma
  \left( v^\mu {y'}^\nu - v^\nu {y'}^\mu \right) \delta^3
  (x-y(\sigma))
  \,.
\end{align}
Here $\sigma$ is an affine parameter describing the string's location
$y^\mu(\sigma,t)$, $v^\mu=(1,\dot y)=dy^\mu/dt$ is the string velocity, and
$H_{\mu\nu\alpha}$ and $A_{\mu\nu}$ are the Kalb-Ramond field strength
and tensor potential, which are a dual representation of $\tA$.
Effectively $\LL_{\mathrm{NG}}$ tracks the effects of the string
tension, which we name $\bar\kappa \pi f_a^2$, stored locally along its
length.  Next, $\LL_{\mathrm{GS}}$ says that the axion angle
propagates under a free wave
equation, as expected for a Goldstone boson, and its decay constant is
$f_a$.  And $\LL_{\mathrm{KR}}$ incorporates the interaction between
strings and axions, also controlled by $f_a$.  The interaction can be
summarized by saying that the string forces $\tA$ to wind by $2\pi$ in
going around the string (in the same sense that the $eJ_\mu A^\mu$
interaction in electrodynamics can be summarized by saying that it
enforces that the electric flux emerging from a charge is $e$).

It should be emphasized that in writing these equations, we are
implicitly assuming a separation scale $\rmin$; at larger
distances from a string $r>\rmin$ we consider $\nabla
\varphi$ energy to be associated with $\tA$; for $r <
\rmin$ the gradient energy is considered as part of the
string tension \cite{Dabholkar:1989ju}, meaning that $\bar\kappa$
incorporates all tension contributions from scales shorter than
$\rmin$.

Any other set of UV physics which reduces to the effective description
of \Eq{Ltot} would present an equally valid way to study this string
network.  Our plan is to find a model without a large scale hierarchy,
such that the IR behavior is also described by \Eq{Ltot} with a large
value for the string tension. Optimally, we want a model which is easy
to simulate on the lattice with a spacing not much smaller than
$\rmin$.  Reading \Eq{LNG} through \Eq{jmn} in order, the
model must have Goldstone bosons with a decay constant $f_a$ and
strings with a large and tunable tension
$T_{\mathrm{str}} = \bar\kappa \pi f_a^2$, with $\bar\kappa \gg 1$.  There
can be other  degrees of freedom, but only if they are very heavy
(with mass $m \sim \rmin^{-1}$), and we will be interested in the
limit that their mass goes to infinity.  Finally, the string must have
the correct Kalb-Ramond charge.  Provided everything is derived from
an action, this will be true if the Goldstone boson mode always winds
by $2\pi$ around a loop which circles a string.

\subsection{The model}

We do this by writing down a model of \textsl{two} scalar fields
$\varphi_1,\varphi_2$, each with a U(1) phase symmetry.  A linear
combination of the phases is gauged; specifically, the fields are
given electrical charges $q_1 \in \mathcal{Z}$ and $q_2 = q_1-1$ under
a single U(1) gauge field.
The orthogonal phase combination represents a global U(1) symmetry
which will give rise to our Goldstone bosons.  The role of the gauge
symmetry will be to attach an abelian-Higgs string onto every global
string, which will enhance the string tension.  The added degrees of
freedom are all massive off the string, achieving our intended
effective description.  The model falls under the general rubric of
``frustrated cosmic strings'' \cite{Hill:1987bw}, but our motivation
and some specifics (particularly our initial conditions) are
different.

Specifically, the Lagrangian is
\begin{align}
  \label{L-2field}
  - \LL(\varphi_1,\varphi_2,A_\mu) & =
  \frac{1}{4e^2} F_{\mu\nu} F^{\mu\nu}
  + \Big| (\partial_\mu -i q_1 A_\mu) \varphi_1 \Big|^2
  + \Big| (\partial_\mu -i q_2 A_\mu) \varphi_2 \Big|^2
  \nonumber \\ & \phantom{=} {} +
  \frac{m_1^2}{8 v_1^2} \Big( 2\varphi_1^* \varphi_1 - v_1^2 \Big)^2
  + \frac{m_2^2}{8 v_2^2} \Big( 2\varphi_2^* \varphi_2 - v_2^2 \Big)^2
  + \frac{\lambda_{12}}{2} \Big(  2\varphi_1^* \varphi_1 - v_1^2 \Big)
  \Big( 2\varphi_2^* \varphi_2 - v_2^2 \Big) \,.
\end{align}
For simplicity we will specialize to the case
\begin{equation}
  \label{masses_equal}
  \lambda_{12}=0, \quad m_1=m_2=\sqrt{e^2(q_1^2 v_1^2 + q_2^2 v_2^2)}
  \equiv m_e \,.
\end{equation}
The model has 6 degrees of freedom; two from each scalar and two from
the gauge boson.  Symmetry breaking,
$\varphi_1 = e^{i\theta_1} v_1\sqrt{2}$ and
$\varphi_2 = e^{i\theta_2} v_2 \sqrt{2}$, spontaneously breaks both
U(1) symmetries and leaves five massive and one massless degrees of
freedom. Specifically, expanding about a vacuum configuration, the
fluctuations and their masses are
\begin{align}
  \label{m-h1}
  v_1 & \to v_1 + h_1 \,, && m = m_1 \\
  \label{m-h2}
  v_2 & \to v_2 + h_2 \,, && m = m_2 \\
  \label{m-A}
  A_i  & \neq 0 \,, && m = \sqrt{e^2(q_1^2 v_1^2 + q_2^2 v_2^2)}\equiv m_e \\
  \label{m-eaten}
  (\theta_1,\theta_2) & \to (\theta_1,\theta_2) + \omega (q_1,q_2) \,, &&
  \mbox{eaten by $A$} \\
  \label{m-goldstone}
  (\theta_1,\theta_2) & \to (\theta_1,\theta_2) +
  \tA \left( \textstyle{ \frac{q_2}{{q_1^2+q_2^2}} ,
    \frac{-q_1}{{q_1^2 + q_2^2}}} \right) && m = 0 \,.
\end{align}
We see that the choices in \Eq{masses_equal} have made all heavy
masses equal.%
\footnote{%
  We set $\lambda_{12}=0$ so that the fluctuations
  in $|\varphi_1|$ and $|\varphi_2|$ are unmixed; our other choices
  ensure that all heavy fields have the same mass.  We could consider
  other cases but we see no advantage in doing so if the goal is to
  implement the model on the lattice.  The lattice spacing is limited
  by the inverse of the heaviest particle mass, while the size of the
  string core and the mass of extra degrees of freedom off the string
  will be set by the inverse of the lightest particle
  mass.  So we get a good continuum limit with the thinnest strings,
  and therefore the best resolution of the network, by having all
  heavy masses equal.}
To clarify, note that a gauge transformation
$A_\mu \to A_\mu + \partial_\mu \omega$ changes
$\theta_1 \to \theta_1 + q_1 \omega$ and
$\theta_2 \to \theta_2 + q_2 \omega$.  Therefore the linear
combination of $\theta_1,\theta_2$ fluctuations with
$\delta \theta_1 \propto q_1$ and $\delta \theta_2 \propto q_2$ is 
precisely the combination which can be shifted into $A^\mu$ by a gauge
change, and is therefore the combination which is ``eaten'' by the
$A$-field to become the third massive degree of freedom.  The
remaining phase difference $q_2 \theta_1 - q_1 \theta_2$ is gauge
invariant,
\begin{equation}
  \label{gauge-invariant}
  q_2 \theta_1 - q_1 \theta_2 \to_{\omega} q_2(\theta_1 + q_1 \omega)
  - q_1(\theta_2 + q_2 \omega)
  = q_2 \theta_1 - q_1 \theta_2 + 0 \omega
\end{equation}
and represents a global, Goldstone-boson mode.

\subsection{The strings}

We initialize $\varphi_1$ and $\varphi_2$ with the \textsl{same}
space-random initial phase, which ensures that all strings will have
\textsl{each} scalar wind by $2\pi$, and the strings will have global
charge $q_1-q_2=1$.  To find the tension of such a string, we write
the \textsl{Ansatz}
\begin{align}
  \label{Ansatz}
  \sqrt{2} \varphi_1(r,\phi) & = e^{i \phi} f_1(r) v_1 \,, \nonumber \\
  \sqrt{2} \varphi_2(r,\phi) & = e^{i \phi} f_2(r) v_2 \,, \nonumber \\
  A_\phi(r) & = \frac{g(r)}{r} \,,
\end{align}
and derive the equations of motion from \Eq{L-2field},
\begin{align}
  \label{g-eq}
  g'' - \frac{g'}{r} & = e^2 v_1^2 f_12 q_1(q_1 g-1)
  + e^2 v_2^2 f_2^2 q_2(q_2 g-1) \,, \\
  \label{f1-eq}
  f_1'' + \frac{f_1'}{r} & = \frac{f_1}{r^2} (1-q_1 g)^2
  + \frac{m^2}{2} f_1(f_1^2-1) \,, \\
  \label{f2-eq}
  f_2'' + \frac{f_2'}{r} & = \frac{f_2}{r^2} (1-q_2 g)^2
  + \frac{m^2}{2} f_2(f_2^2-1) \,.
\end{align}
Here $f_1,f_2$ represent the progress of the two scalar fields towards
their large-radius asymptotic vacuum values, while $2\pi g(r)$ is the
magnetic flux enclosed by a loop at radius $r$, which trends at large
$r$ towards the total enclosed magnetic flux.
The large-$r$ behavior is well behaved only if $f_1\to 1$,
$f_2 \to 1$, and
\begin{align}
  \label{g-limit}
  \lim_{r\to \infty}  g(r) = \frac{q_1 v_1^2 + q_2 v_2^2}
      {q_1^2 v_1^2 + q_2^2 v_2^2}
      \quad = \frac{1}{2\pi}\mbox{(enclosed magnetic flux)}\,.
\end{align}
The magnetic flux is therefore a compromise between the value
$1/q_1$, which cancels large-distance gradient energies for the first
field, and $1/q_2$, which cancels large-distance gradient energies for
the second field.  The gradient energy at large distance is given by
\begin{align}
  \label{tension-fa}
  T_{\mathrm{str}} & \simeq 2\pi \int r \: dr \left( |D_\phi \varphi_1|^2 +
  |D_\phi \varphi_2|^2 \right) \nonumber \\
  & \simeq \pi \int r \: dr \left( \frac{v_1^2}{r^2}
  \Big( 1 - q_1 g \Big)^2 + \frac{v_2^2}{r^2}
  \Big( 1 - q_2 g \Big)^2 \right)
  \nonumber \\
  & \simeq \pi \int \frac{dr}{r} \:
  \frac{v_1^2 v_2^2}{q_1^2 v_1^2 + q_2^2 v_2^2} \,.
\end{align}
Comparing \Eq{Tension}, \Eq{kappa} with \Eq{tension-fa}, we identify
the Goldstone-mode decay constant as
\begin{equation}
  \label{fa-is}
  f_a^2 = \frac{v_1^2 v_2^2}{q_1^2 v_1^2 + q_2^2 v_2^2} \,.
\end{equation}

\begin{figure}[htb]
  \hfill \includegraphics[width = 0.4\textwidth]
    {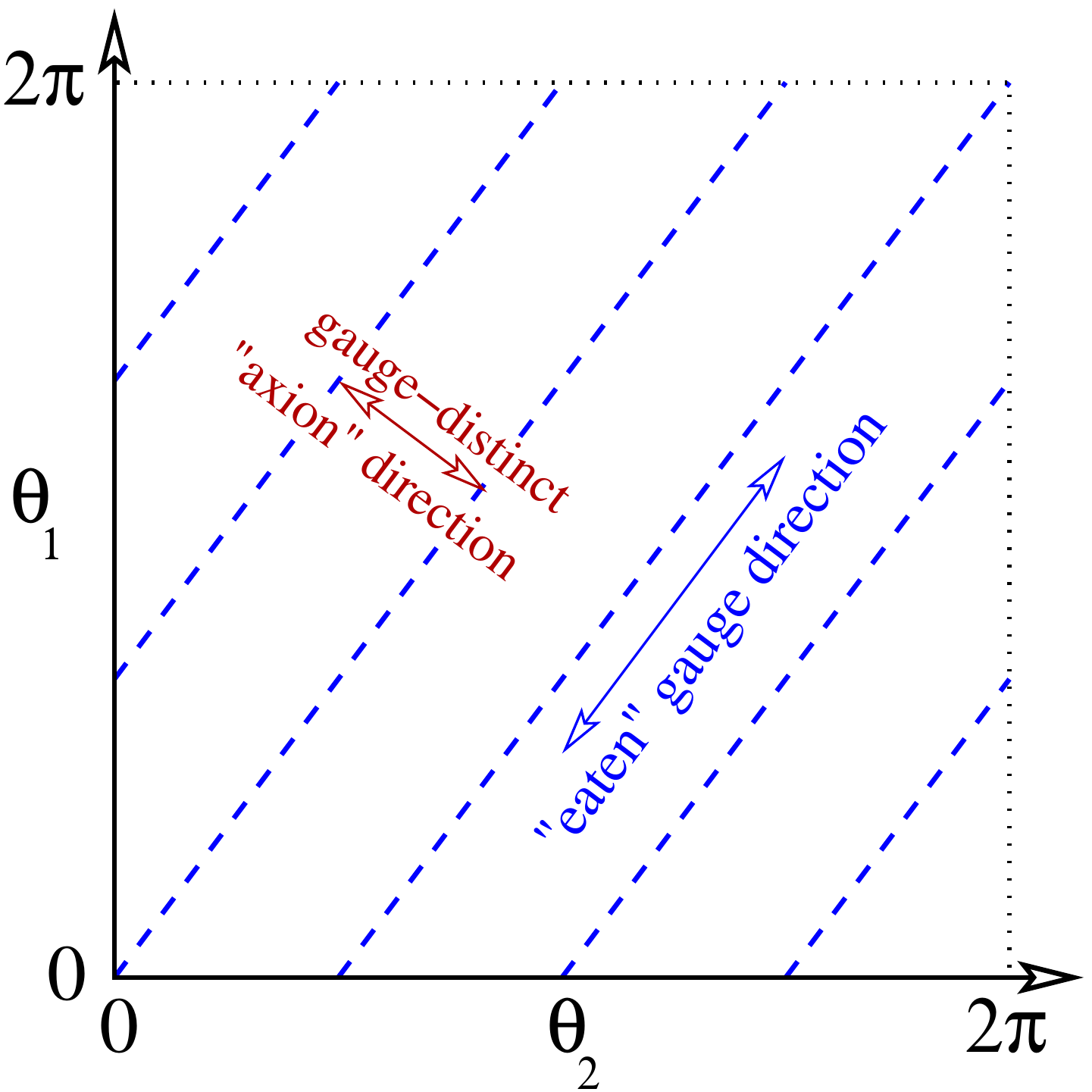}
  \hfill $\phantom{.}$
  \caption{\label{fig:stripes}
  Space of $\varphi_1,\varphi_2$ phases $(\theta_1,\theta_2)$ for the
  case $(q_1,q_2)=(4,3)$.  The dashed (blue) line indicates phase
  pairs which are equivalent under gauge transformations.  An
  appropriate vector potential can cancel any gradient energy in the
  direction of the dashed lines.}
\end{figure}

\begin{figure}[htb]
  \hfill \includegraphics[width = 0.6\textwidth]{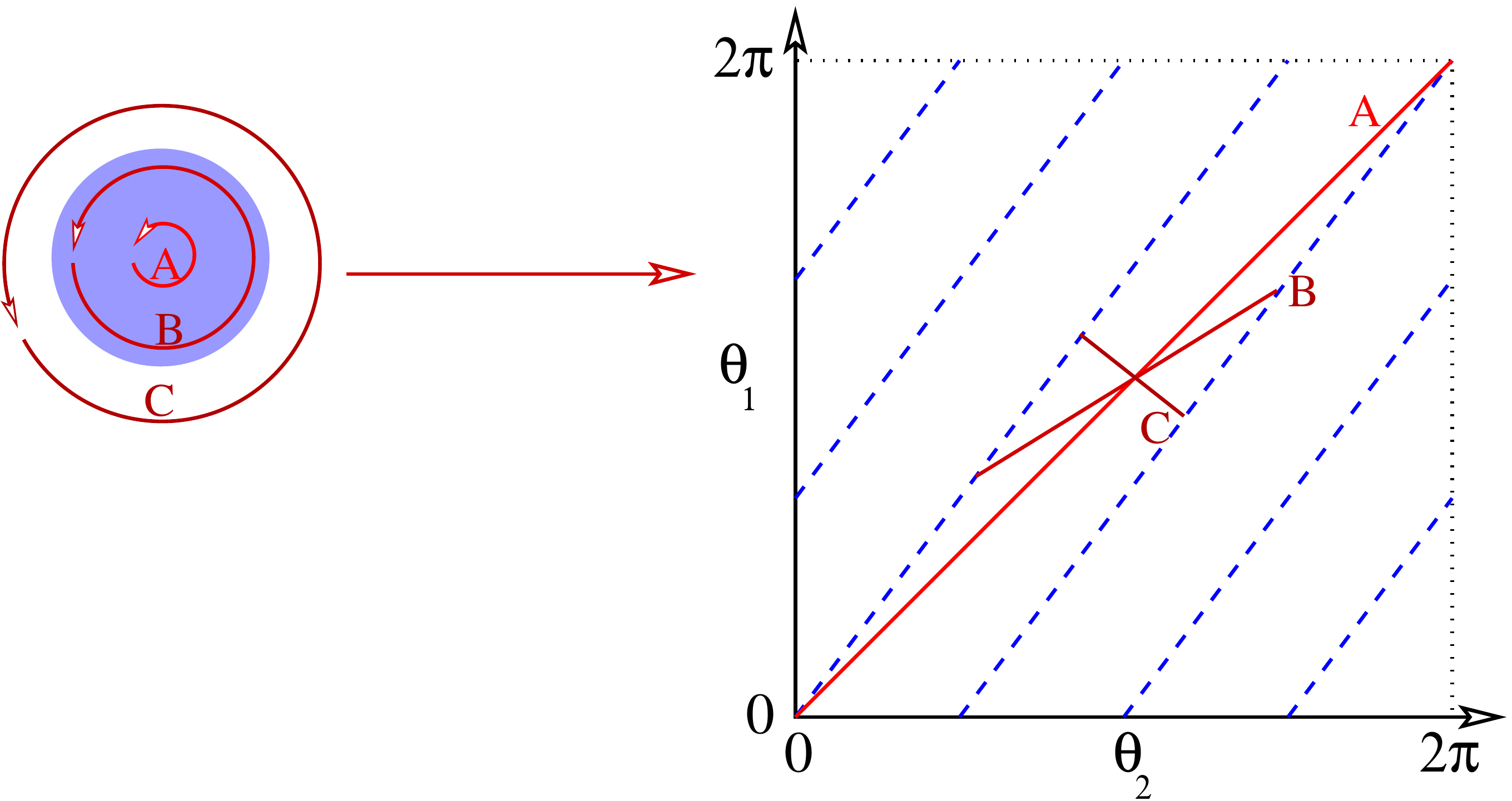}
   \hfill $\phantom{.}$
  \caption{\label{fig:strings}
  Left:  cross-section of a string, showing the magnetic field
  strength ``bundle'' and three possible loops one can take around the
  center of the string.  Right:  path through $(\theta_1,\theta_2)$
  space taken along each loop.  As more magnetic flux is enclosed, the
  component of $(\Delta \theta_1,\Delta \theta_2)$ along the
  gauge-direction is canceled, but the component in the ``global''
  direction is not.}
\end{figure}

For a more intuitive explanation, consider Figure \ref{fig:stripes}.
It shows the set of possible phases $(\theta_1,\theta_2)$ for the two
scalar fields, in the case $(q_1,q_2) = (4,3)$.  The figure includes
a dotted line to indicate which
phase choices are gauge-equivalent.  Moving along the dotted line
corresponds to changing the gauge, or moving through space along a
gauge field; a vector potential of the right size can cancel a gradient
energy along this field direction.  The orthogonal direction, which
is unaffected by a gauge field, is the global (axion) field
direction.  A change in this direction from one blue dotted line to
the next represents a full $2\pi$ rotation in the (axial) Goldstone
direction, which explains the value of $f_a$ found in
\Eq{fa-is}.  Figure \ref{fig:strings} then shows how each field varies
around a string.  As we consider loops farther and farther from the
string's center, more and more flux is enclosed, so more and more of
the gradients along the blue-dotted direction are canceled by the
$A_\phi$ field.  For the innermost loop there is no
enclosed flux, and the gradient energy is given by the distance
between the point $(\theta_1,\theta_2)=(0,0)$ to the point
$(2\pi,2\pi)$.  For a loop enclosing the entire flux, all gradient
energy arising from the gauge-direction is canceled, almost but not
fully removing the gradient energy.  Only the gradient energy arising
from the shortest path from one blue-dotted line to the next cannot be
compensated.  This represents the residual global charge of the
string.  This path length is $2\pi f_a$.

Now let us estimate the effective value of $\bar\kappa$, the added
contribution to the string tension in units of the long-distance
Goldstone-mode contribution.  The energy of the string's core is the
energy of an abelian Higgs string with $m_h = m_e$ and with
$f^2 = v_1^2 + v_2^2$, which is
\begin{equation}
  \label{abelian-tension}
  T_{\mathrm{str,abelian}} \simeq \pi ( v_1^2 + v_2^2 ) \,.
\end{equation}
The value of $\bar\kappa$ is therefore
\begin{equation}
  \bar\kappa = \frac{T_{\mathrm{str}}}{\pi f_a^2}
  \simeq \frac{v_1^2 + v_2^2}
  { \frac{v_1^2 v_2^2}{q_1^2 v_1^2 + q_2^2v_2^2}}
  = \frac{(v_1^2+v_2^2)(q_1^2 v_1^2+q_2^2 v_2^2)}{v_1^2 v_2^2}
  \quad
  \longrightarrow_{v_1=v_2}
  \quad
  2(q_1^2 + q_2^2) \,.
\label{kappa-eff}
\end{equation}
Detailed calculations show that this is indeed the added tension.
The full value of $\kappa$ is $\kappa = \bar\kappa + \ln(m/H)$ where
$m,H$ are the values actually used in the numerical simulation;
typically $m/H \sim 1000$.  So choosing $q_1=4$ and $v_1=v_2$ gives
$\bar\kappa = 50$ and $\kappa = 57$, in the middle of the physically
interesting range.

\begin{figure}[htb]
  \hfill \includegraphics[width = 0.5\textwidth,bb=27 9 416 331]{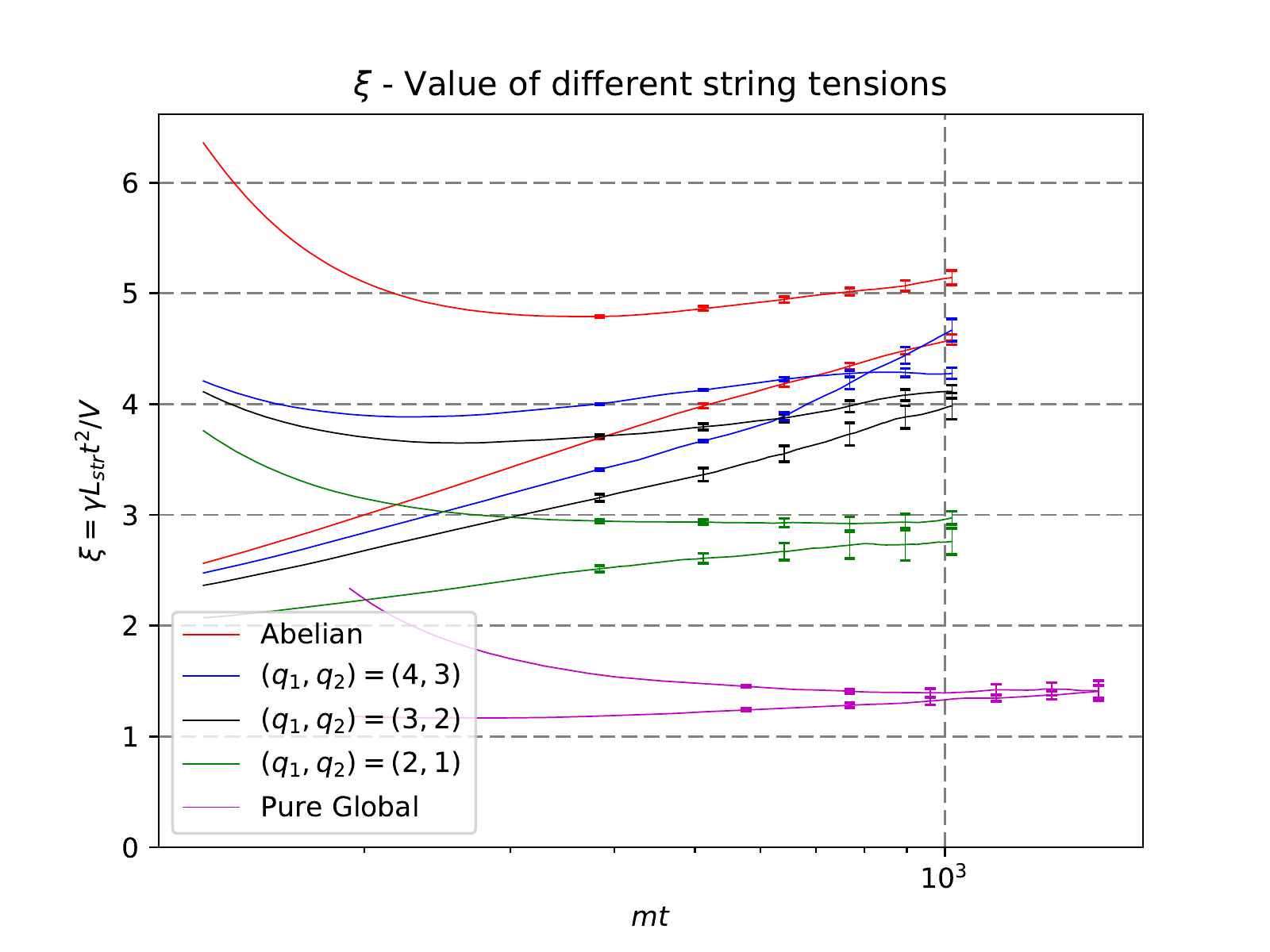}
  \hfill $\phantom{.}$
  \caption{\label{fig:length}
    Network density for different string tensions. The
    falling curves represent the overdense networks, while
    the rising curves represent the underdense
    networks.} 
\end{figure}

Figure \ref{fig:length} shows how the density of the string network
depends on the string tension.  The network density is expressed in
terms of the dimensionless scaling variable $\xi$,
\begin{equation}
  \label{xi-def}
  L_{\mathrm{sep}}^{-2}
  \equiv V^{-1} \int_{\mathrm{all\;string}} \gamma dl \,, \qquad
  \xi \equiv \frac{t^2}{4 L^2} \,.
\end{equation}
We see that it increases by over a factor of 3 as one goes from
scalar-only ($\kappa \simeq 7$) to high-tension ($\kappa \simeq 57$)
simulations.  It is not clear that our lattices are large enough to
see the onset of scaling behavior for the highest-tension networks we
studied.  We certainly don't see scaling behavior for the abelian
Higgs simulations, but this is another story.

\subsection{Numerical implementation}

I will not insult this audience by explaining how to implement a
bosonic U(1) theory on the lattice.  I use the noncompact formulation
of U(1) and a next-nearest neighbor $(a^2)$ improved action.  There
are no issues of renormalization of parameters because we are studying
classical field theory.  A
subtlety in implementing electric fields with improvement is handled
as in \cite{Moore:1996wn}.  We pause only to mention one subtlety in
how we implement $\chi (1-\cos\tA)$.  The function $\cos\tA$, with
$\tA = q_2 \theta_1 - q_1 \theta_2$, is highly singular near the
string core.  Such singular behavior creates problems under space
discretization, so we have to ``round off'' the behavior inside the
string core, with the substitution
\begin{align}
\label{chilatt}
  \chi(t) \Big( 1 - \cos \tA \Big)
  & \Rightarrow \chi(t)
  F(2\varphi_1^*\varphi_1) F(2\varphi_2^* \varphi_2)
  \left( 1 - \cos \Big( q_2 \argphi_1 - q_1 \argphi_2 \Big) \right)  \,,
  \\
\label{Flatt}
  F(r) & \equiv \left\{ \begin{array}{ll}
    \frac{25}{16} r \left( \frac{8}{5} - r \right)\,, \quad
    & r< \frac 45 \,, \\
    1\,, & r > \frac 45 \,. \end{array} \right.
\end{align}
The smoothing function $F(r)$ is chosen such that
$F(1)=1$, $F'(1)=0$, $F(0)=0$, and $F'(r)$ is continuous.
This modification only affects the dynamics inside string cores, but
the $\chi(t) (1-\cos\tA)$ term is much smaller than the other
potential terms there.  Indeed, this term is everywhere small and it
is only important because it operates over much of the lattice volume,
while the radial potential has an effect only in the tiny fraction of
the lattice corresponding to string cores.

\section{Results}

We studied this model \cite{axion4} on $2048^3$ lattices using a
single compute node containing two Xeon Phi (KNL) processors.  After
systematic studies of lattice spacing and continuum limits, we also
studied how the network evolution and the axion production depend on
the string tension.

\begin{figure}[htb]
  \includegraphics[width=0.46\textwidth,bb=46 160 550 592]{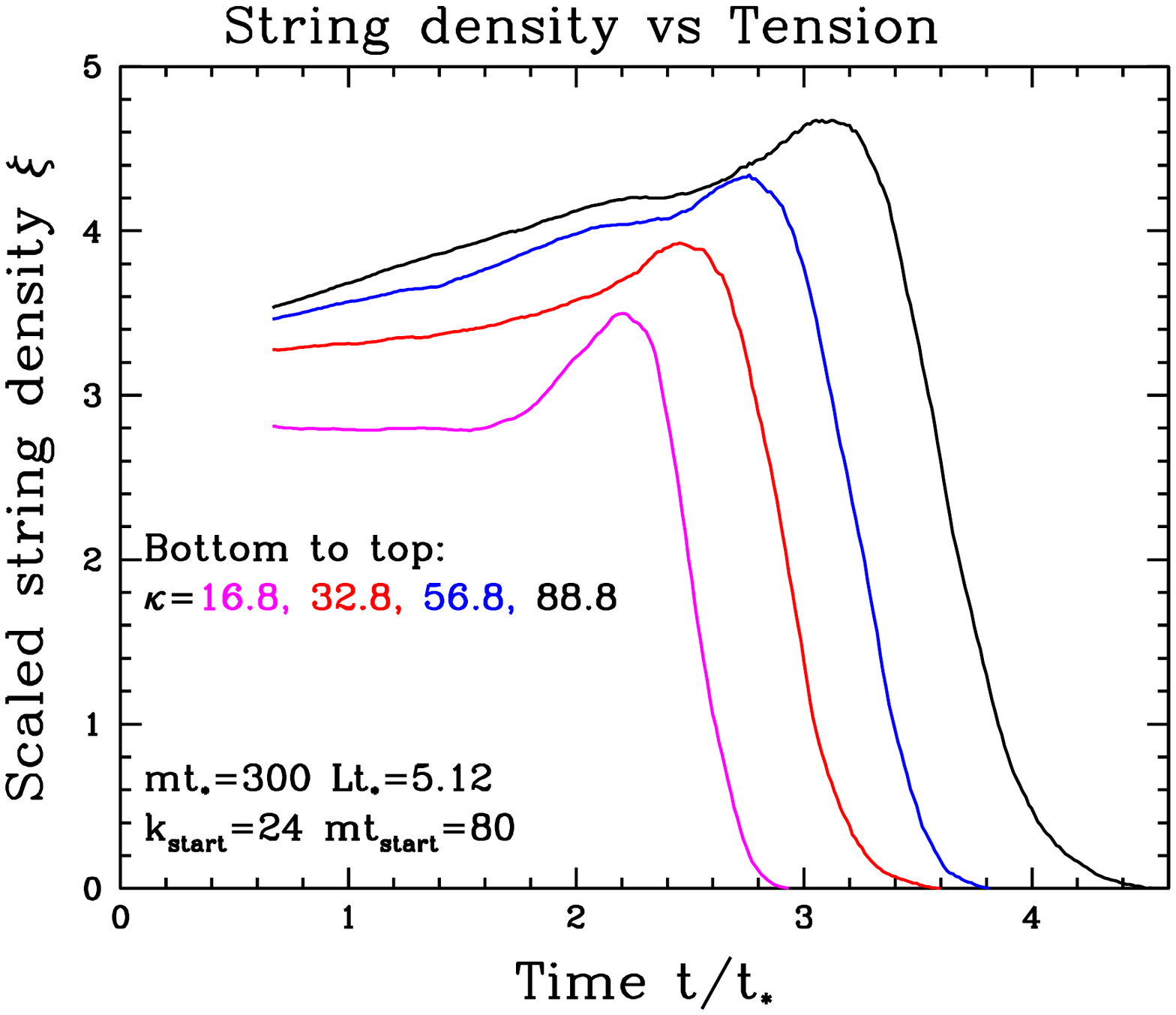}
  \hfill
  \includegraphics[width=0.46\textwidth,bb=21 143 558 591]{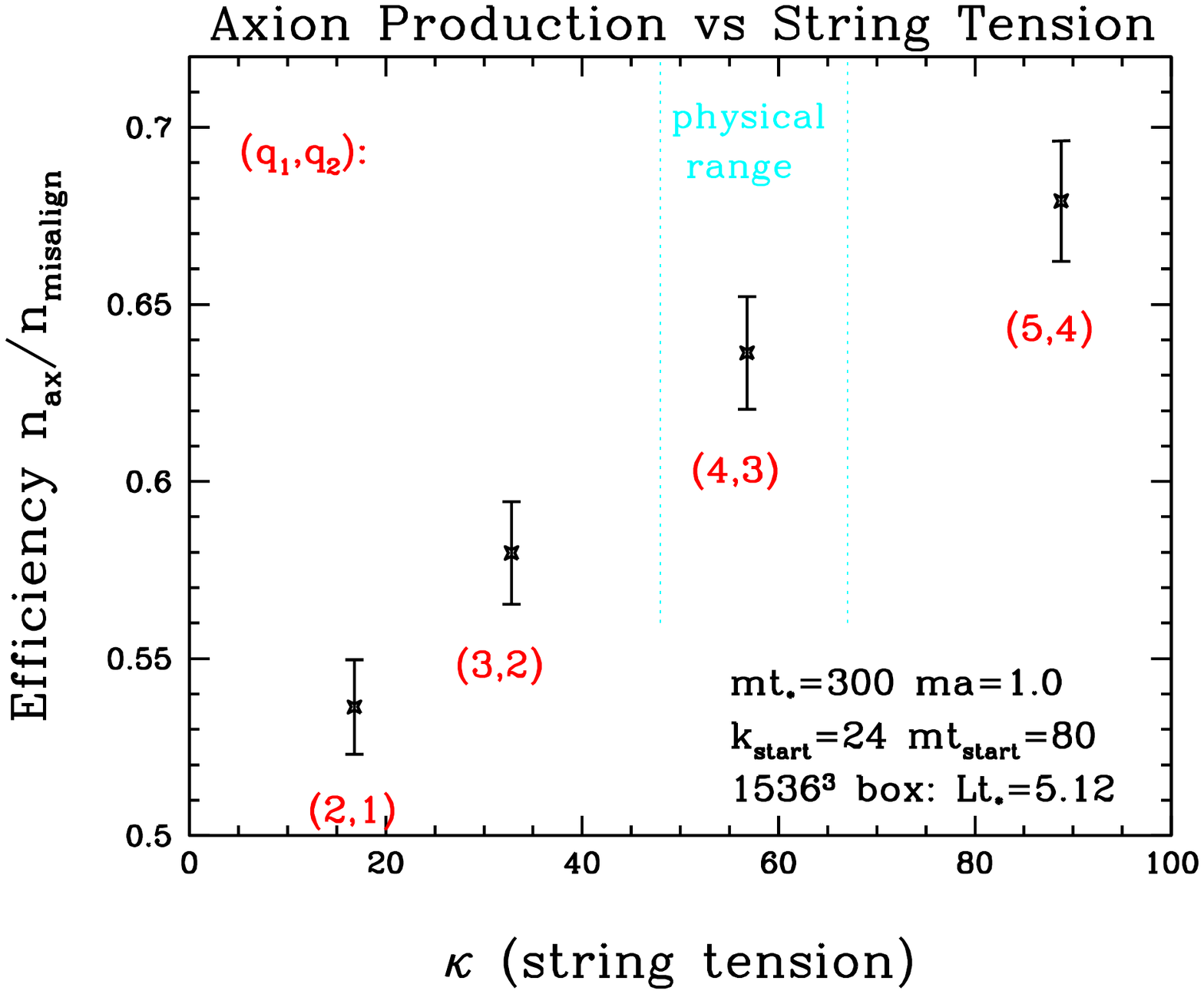}
  \caption{\label{fig:kappa}
    Left:  string density as a function of time for different $\kappa$
    values.  The higher the string tension, the longer the strings
    persist.  Right:  axion production efficiency as a function of
    $\kappa$.
  }
\end{figure}

The most notable result is that the axion production is actually
\textsl{smaller} for the case with strings than the angle-average of
the ``misalignment'' mechanism of \Eq{EOMax}.  This contradicts the
``conventional wisdom'' \cite{Hiramatsu:2012gg} that axion production
should be the sum of a misalignment contribution, a string
contribution, and a wall contribution.  We claim that this view double
counts; the energy in domain walls \textsl{is} the energy of field
misalignment, from values $\tA \sim \pi$.  This energy represents most
of the potential axion production from misalignment.  But the energy
in the wall network is mostly absorbed by the strings when the walls
pull on the strings, giving their energy to string velocity.  Then the
strings chop up into loops, which appear to produce axions quite
inefficiently.

Combining our numbers with expressions for the energy budget $g_*$ and
topological susceptibility from \cite{Borsanyi:2016ksw}, and the
observed dark matter density, we find
$m_a = 26.2 \pm 3.4 \, \mu\mathrm{eV}$.

\section{Topology: temperature range}

We have emphasized that an input value for $\chi(T)$ from the lattice
is essential to deriving these results.  But we don't need $\chi(T)$
at all temperatures; some are more important than others.  We will
extend the results of \cite{axion4} by investigating over what
temperature range the topological susceptibility is really needed.

At sufficiently high temperature $\chi(T)$ is small.  Therefore
$\chi(T)(1-\cos\tA)$ plays little role in the field dynamics.  Its
importance is controlled by the combination $m_a t$ which rises with
time as approximately $t^{5.8}$.  Therefore at high temperature, large
errors in $\chi(T)$, or no value at all, is not a problem.  To study
this, we replace $\chi(T)$ with the following ``chopped'' form:
\begin{equation}
  \label{cut1}
  \chi(T) \to \left\{ \begin{array}{ll}
    \chi(T) & T < \Tchop \\
    \chi(\Tchop) & T > \Tchop \,. \\
    \end{array} \right.
\end{equation}
Technically it is $t^2 \chi(T)$ we chop in this way, because of the
$t^2$ factor in, eg, \Eq{EOMax}.  We then study the axion production
as a function of the temperature $\Tchop$.  When $\nax$
ceases to depend on $\Tchop$, we know that $\chi(\Tchop)$ is not
relevant.  But so long as the result with $\Tchop$ is different than
the unchopped limit, we need $\chi(T)$ at that temperature.

\begin{figure}[thb] 
  \centering
  \includegraphics[width=8cm,clip]{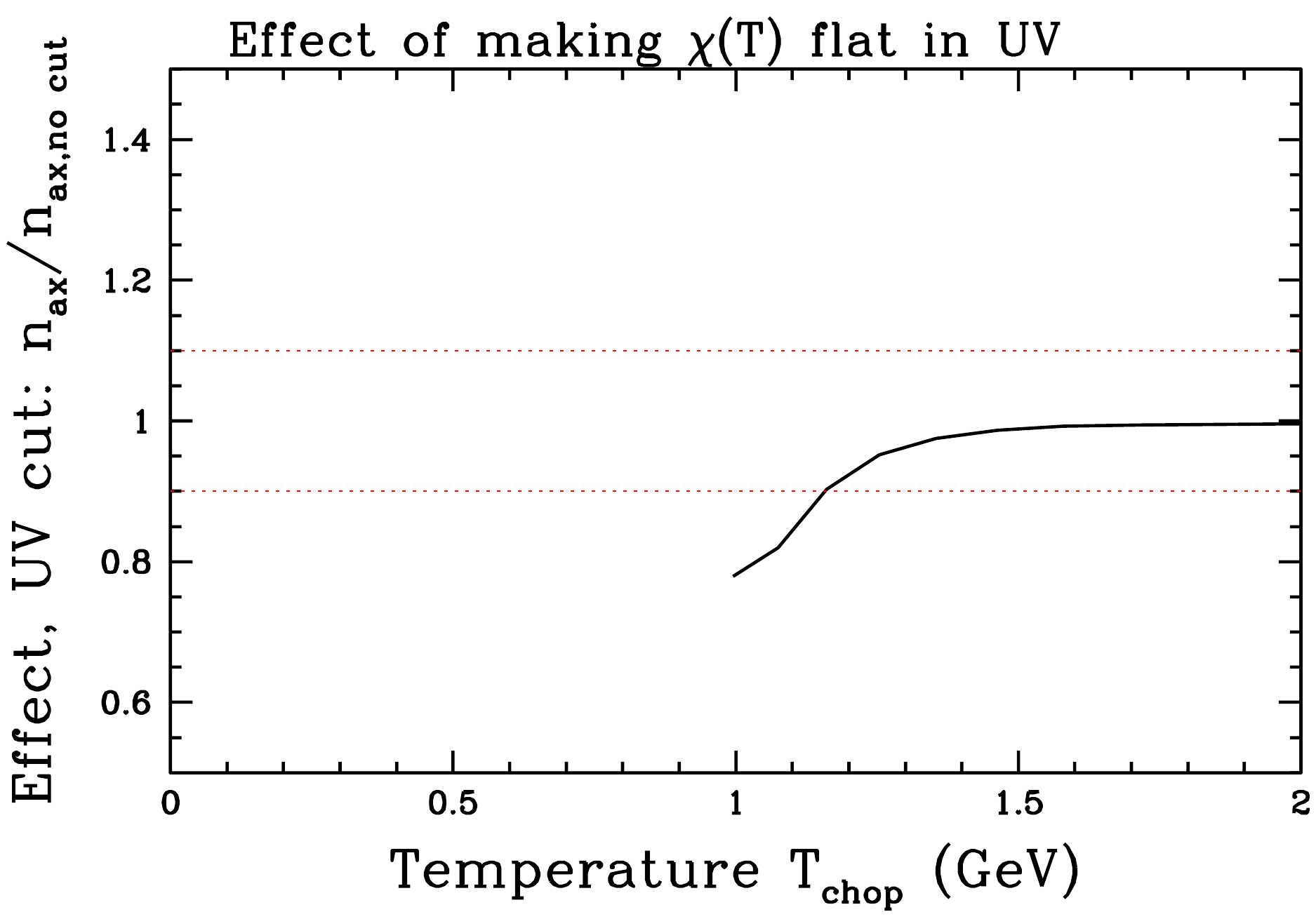}
  \caption{Effect of replacing $\chi(T)$ with a result which is flat
    in the UV, to diagnose at what scale we need to compute $\chi(T)$.}
  \label{fig:tmax}
\end{figure}

We see from Fig.\ref{fig:tmax} that above about 1150 MeV, it makes little
difference if we change $\chi(T)$.  But below this value, we are quite
sensitive.  So it is necessary to determine $\chi(T)$ all the way up
to 1150 MeV.

On the IR side, perhaps surprisingly, there is also a range of
temperatures where the susceptibility is not important.  That is
because the strings and walls are gone and the $\tA$ value is small
and oscillating rapidly, with $m_a \gg H$.  Then the axion number is
an approximate adiabatic invariant, which reacts smoothly to Hubble
expansion and mass shifts and does not feel the anharmonicity of the
potential, $1-\cos\tA \simeq \tA^2/2$.  All that matters is the final
value of $\chi(T\to 0)$, which is well known.  To find out where this
temperature range starts, we make a similar modification:
\begin{equation}
  \label{cut2}
  \chi(T) \to \left\{ \begin{array}{ll}
    \chi(T) & T > \Tchop \\
    \chi(\Tchop) & T < \Tchop \,. \\
    \end{array} \right.
\end{equation}
That is, we freeze $\chi(T)$ (really, $t^2 \chi(T)$) from rising after
some point, and see if that changes the axion number produced.

\begin{figure}[thb] 
  \centering
  \includegraphics[width=8cm,clip]{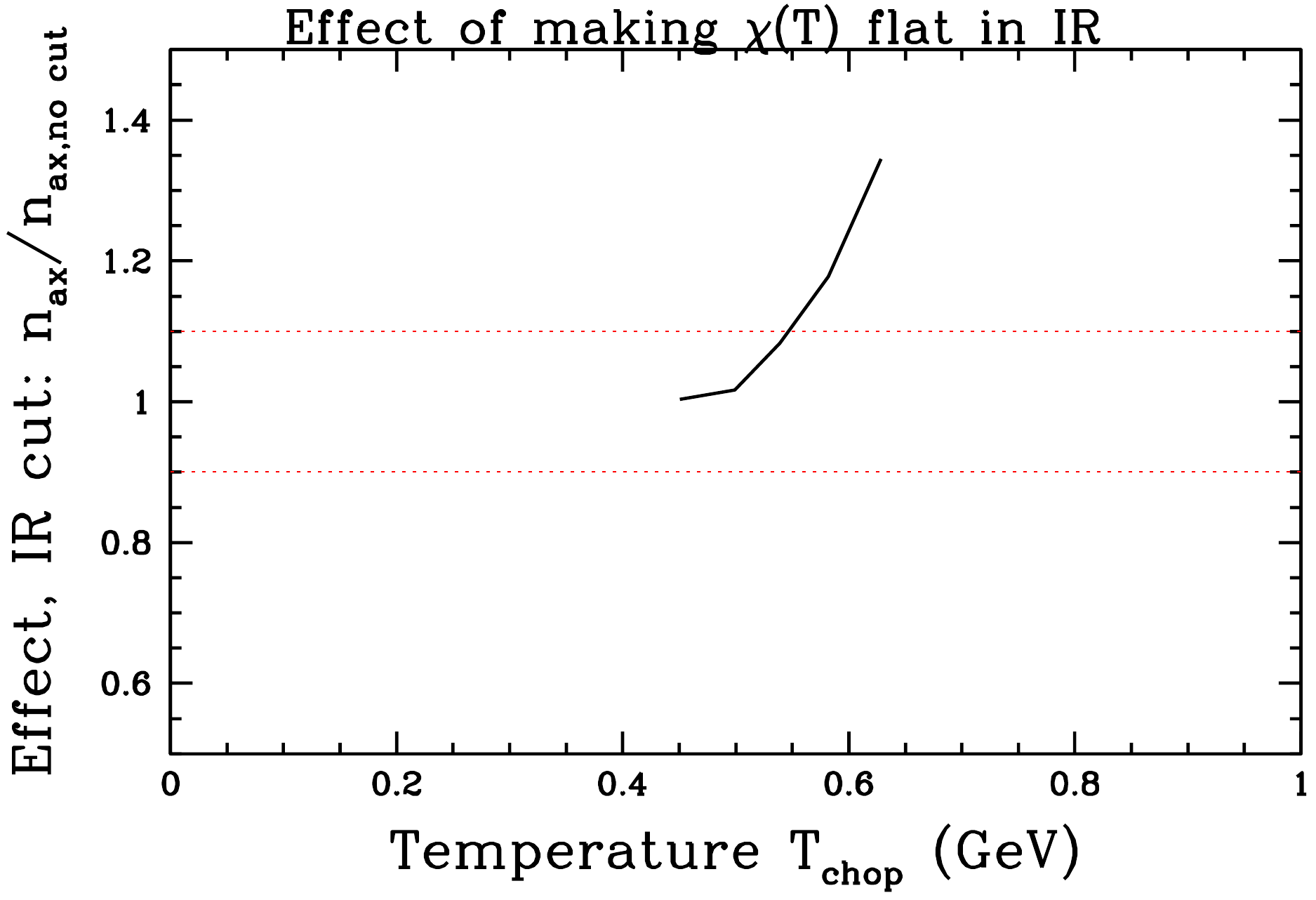}
  \caption{Effect of replacing $\chi(T)$ with a result which is flat
    in the IR, to diagnose at what temperature we first need to know
    $\chi(T)$.}
  \label{fig:tmin}
\end{figure}

The results are shown in Fig.~\ref{fig:tmin}. They indicate that the
susceptibility is irrelevant below about 550 MeV; the behavior above
that temperature is important.  Therefore the axion dynamics is
sensitive to $\chi(T)$ in the range 550 to 1150 MeV; outside of that
range it is not.

\section{Conclusions}

The axion is a well motivated hypothetical particle, because it might
explain two mysteries -- the P and T symmetry of QCD, and the nature
of the Dark Matter -- with a single mechanism and particle.
The physics of the axion in cosmology is rich, governed by a network
of string defects which are swept together when domain walls form.
Its explication requires two new pieces of physics.  We need to
understand this network evolution and axion production better.  And we
need to know the QCD topological susceptibility as a function of
temperature, $\chi(T)$, because it sets the tension of the domain
walls and controls the physics which destroys the string network.  I
have presented the latest details on the network evolution,
introducing a new technique which allows simulation of high-tension
strings without excessive numerical resources.  The results indicate
rather inefficient axion production and therefore a rather light axion
compared to previous studies, with
$m_a = 26.2 \pm 3.4 \,\mu \mathrm{eV}$.

It remains to form a consensus in the lattice community that the
topological susceptibility is well measured.  My work indicates that
the susceptibility is needed in a temperature range from 550 to 1150
MeV.  Below this range the evolution is adiabatic.  Above this range
the susceptibility does not yet play a role in axion dynamics.
Clearly it is challenging to determine the susceptibility at such high
temperatures.  But I am confident this is a challenge which the
lattice community will accept with gusto.

\section*{Acknowledgments}

I am indebted to close collaboration with Vincent Klaer and Leesa
Fleury, and to useful conversations with Mark Hindmarsh.  This work
has been supported both by the TU Darmstadt's Institut f\"ur
Kernphysik and the GSI Helmholtzzentrum, and by the Department of
Physics at McGill University and the Canadian National Science and
Engineering Research Council (NSERC).

\clearpage
\bibliography{refs}

\end{document}